\documentstyle[12pt]{article}

\begin{document}
Cavendish preprint HEP 94/3 \\
12th April 1994 \\
\vspace{1.0cm}
\begin{center}
{\huge \bf CONFINEMENT AND THE PION NUCLEON SIGMA TERM \\}
\vspace{3ex}
\vspace{3ex}
{\large \bf S. D. Bass \\}
\vspace{3ex}
{\it HEP Group, Cavendish Laboratory, \\
University of Cambridge, Madingley Road, Cambridge, CB3 0HE,
England \\}
\vspace{3ex}

\vspace{3ex}
{\large \bf ABSTRACT \\}
\end{center}
\vspace{3ex}
{Gribov's theory of confinement offers a simple explanation of the value of
the pion nucleon sigma term.
There is no need to invoke a large strange quark component in the nucleon.}

\vspace{1.5cm}

It has long been thought that there is a discrepancy between the value of
the pion nucleon sigma term $\sigma_{\pi N}$ measured in $\pi$ $N$ scattering
and the theoretical prediction of $\sigma_{\pi N}$ from hadron spectroscopy
(see eg. [1] and references therein).
In this paper I explain why this ``discrepancy" is a natural consequence of
Gribov's mechanism for confinement [2, 3].

It will be helpful to first review the theory of $\sigma_{\pi N}$ in the usual
theory of pions and chiral symmetry [4].
The pion nucleon sigma term is a measure of chiral symmetry breaking in the
nucleon.
The QCD Lagrangian exhibits exact chiral symmetry for massless quarks.
Since there are no parity doublets in the hadron spectrum we know that
the chiral symmetry must be spontaneously broken, whence Goldstone's theorem
tells us to expect a zero mass boson.
In the real world the light quarks have a small mass, which breaks the exact
chiral symmetry.
We use $H_m = {\hat m} {\overline q}q$ to denote the chiral symmetry breaking
term in the QCD Lagrangian
where ${\overline q}q = {\overline u}u + {\overline d}d$ and ${\hat m}$ is the
mean light ``current-quark" mass.
If we assume that hadronic physics changes continuously (with no phase
transition)
as we vary the light quark mass from zero to ${\hat m}$
then the chiral Goldstone state acquires a small mass. It is identified with
the physical pion.
The theory of chiral symmetry gives a relation between the value of
${\hat m}$ and the pion mass, viz.
$$
m_{\pi}^2 = {\hat m} \Biggl( {- <vac| {\overline q}q |vac> \over f_{\pi}^2 }
\Biggr)
\eqno(1)
$$
which we shall need later in our discussion.
The sigma term is formally defined as
$$
\sigma_{\pi N} = {1 \over 3} \sum_{i=1}^3
<N| \Biggl[ Q_5^i, \Biggl[ Q_5^i, H_m \Biggr] \Biggr] | N>
\eqno(2)
$$
where $Q_5^i$ is the axial charge.
After we use the QCD equations of motion to evaluate the commutators in equ.(2)
$\sigma_{\pi N}$
becomes
$$
\sigma_{\pi N} = \int d^3 x {\hat m} <N| {\overline q}q |N>
\eqno(3)
$$
The value of $\sigma_{\pi N}$ is measured in $\pi$ $N$ scattering to be
$\sigma_{\pi N} \simeq 45$MeV [5].
In renormalised QCD $[m {\overline q}q]$ is scale invariant. For this reason
it is commonly assumed that vacuum polarisation in QCD
does not play an important role in the physics of chiral symmetry.
Using the QCD sum-rule determination of $- <vac| {\overline q}q |vac>$ one
finds that ${\hat m} =6$MeV at a scale $\mu^2 = 1$GeV$^2$ [6].

The determination of $\sigma_{\pi N}$ from hadron spectroscopy goes as follows.
The mass degeneracy in the baryon octet is broken by the finite quark masses.
If we make a leading order analytic (linear) expansion in the quark mass and
assume
that there is a negligible strange quark component $<N| {\overline s}s |N> =0$
in the nucleon,
then it is easy to show that [6]
$$
\sigma_{\pi N}^{th} = {3 (M_{\Xi} - M_{\Lambda}) \over ({m_s \over m_q} -1) }
\eqno(4)
$$
where $m_q$ and $m_s$ are the running light quark and strange quark masses
respectively.
If we identify
$m_q = {\hat m}$
(whence ${m_s \over m_q} = 25$), then we find the
familiar
prediction of hadron spectroscopy $\sigma_{\pi N}^{th} = 25$MeV.
The difference between this theoretical prediction and the value of
$\sigma_{\pi N}$ measured
in $\pi$ $N$ scattering is the sigma term ``discrepancy".
It has led to suggestions (see [1]) that there might be a large strange quark
component $<N| {\overline s}s |N>$ in the nucleon.
If this were true then a significant fraction of the nucleon's mass would be
due
to strange quarks --
in contradiction with the quark model.

The derivation of equs.(1-4) assumed that hadronic physics changes continuously
(with no phase transition) as we take $m_q \rightarrow 0$.
At this point we have to be careful.
Even in QED we know that the theory of the electron differs from the theory
with
a zero mass gap.
The Born level cross section for $e^+ e^-$ production when a hard (large $Q^2$)
transverse photon scatters from a soft (small $-p^2$)
longitudinal photon is non-vanishing when we let $-p^2 \rightarrow 0$ in QED
with a zero mass gap [7].
This is in contrast to the familiar physical situation where the electron has a
finite
mass and this cross-section vanishes as $-p^2 \rightarrow 0$.
As we take $m_e \rightarrow 0$
the vacuum in QED becomes strongly polarised:
the perturbation theory expression for the vacuum polarisation $\Pi (q^2)$
diverges logarithmically.
This suggests that the vacuum state for QED with a zero mass gap is not
perturbative:
it exists in a different phase of the theory [8].
In QCD there is every reason to expect vacuum polarisation to play an
important role
in the physics of light quarks
since the QCD dynamics at strong coupling must
spontaneously break the chiral symmetry of the classical theory.

We now present a simple explanation of the sigma term ``discrepancy" in terms
of Gribov's theory of confinement [2,3].
Gribov's idea is that QCD becomes super-critical at some finite
$\alpha_s^c \sim 0.6$,
at which point the energy level of a quark in a background colour field falls
below the Fermi surface of the perturbative vacuum.
The quark then becomes a resonance and is not seen as a free particle.

The Gribov theory is obtained via the simultaneous solution of the
Schwinger-Dyson equation for the quark propagator and the Bethe-Salpeter
equation for meson bound states.
At a critical coupling
$\alpha_s^c \sim 0.6$ the theory exhibits a rich sequence of phase transitions.
There appear multiple solutions to the quark propagator equation which
correspond to new states in the light quark vacuum.
These new vacuum states
correspond to quasi-particle excitations with both positive and negative
masses [2].
Each phase transition leading to new states in the vacuum is characterised
by a critical mass $m_P^c$.
Provided that the running quark mass $m_P$ at the critical scale $\lambda$
(where $\alpha_s = \alpha_s^c$)
is less than the critical mass $m_P^c \ll \lambda$ one finds that the solution
of the Schwinger-Dyson equation
in the new vacuum states
matches onto the solution of the perturbative renormalisation group equation,
which describes
the physics at small coupling.
If this condition is satisfied then the new vacuum states yield physical
excitations in the theory.
(For technical details and a derivation of these results see [2].)

Each transition is characterised by a pseudo-scalar Goldstone bound state.
(The appearance of the new vacuum states spontaneously breaks the chiral
symmetry.)
The wavefunctions of the physical
Goldstone states are found by perturbing the solution of the Bethe-Salpeter
equation
about the critical mass $m_P^c$.
The mass of the Goldstone state associated with any given phase transition
is determined by $m_P$ and the value of $m_P^c(\lambda)$ for that transition.
One finds that
the Goldstone meson mass is [2]
$$
\mu_{\pi}^2 = \kappa^2 (m_P^c - m_P)
\eqno(5)
$$
where $\kappa^2$ is a positive constant.
The fact that we see only one Goldstone pion state in the physical spectrum
tells us that
the light quark mass lies somewhere
between the critical mass for the first and the second vacuum transitions.

We compare equs. (5) and (1),
whence it follows that
the ``current-quark" mass in pion physics (and in particular equ.(3)) is
$$
{\hat m} = m_P^c - m_P
\eqno(6)
$$
rather than $m_P$.
The QCD dynamics which lead to the chiral phase transition
tell us that chiral perturbation theory (in pion physics) is really an
expansion about
$(m_P^c - m_P) =0$ rather than about $m_P =0$.
(The usual formulation of chiral symmetry {\it assumes} that $m_P^c =0$, which
need not be the case.)
As we decrease $m_P \rightarrow 0$ the mass of the {\it physical} pion
increases in
the Gribov theory.
When $m_P$ coincides with the critical mass for the n$^{th}$ transition
($n \geq 2$)
a new Goldstone state appears in the hadron spectrum with zero mass.
The mass of this state increases as we decrease $m_P$ further.
At $m_P = 0$
one finds an infinite number of pseudo-scalar Goldstone states with small
masses
$\mu_{\pi,n}^2$, where $\mu_{\pi,n}^2 \geq \mu_{\pi,n+1}^2 \rightarrow 0$
as $n \rightarrow \infty$ [2].

If we compare equs. (1) and (6), it follows that
$$
\kappa^2 = \Biggl[ { - <vac| {\overline q}q |vac> \over f_{\pi}^2 }
\Biggr]_{\mu^2 = \lambda}
\eqno(7)
$$
in the usual theory of pion physics [4].
Using the scale invariance of $[m {\overline q}q]$, chiral perturbation theory
tells us that
$$
m_P^c - m_P = 6 MeV
\eqno(8)
$$
{\it after} we evolve to $\mu^2= 1$GeV$^2$.

Whilst the ``current-quark" mass in pion physics is ${\hat m}=m_P^c-m_P$
(equ.(6)),
the mass in the renormalised QCD Hamiltonian is (of course) the running quark
mass $m_P$.
The mass degeneracy of physical non-Goldstone states (like the baryon octet)
is broken by the finite values of the running masses $m_P$.
The baryon mass $M_B$
is sensitive to two effects as we vary the light quark
mass between zero and the critical mass $m_P^c$ for the first transition.
Firstly, the mass of the constituent quark quasi-particle excitation
increases
with $m_P$.
When $m_P$ is less than the critical mass for the n$^{th}$ transition the
baryon mass also receives ``pion" corrections associated with the n$^{th}$
Goldstone state.
The leading ``pion" correction to $M_B$ in improved chiral perturbation theory
is proportional to $\mu_{\pi}^2 \ln \mu_{\pi}^2$ [6],
where $\mu_{\pi}^2$ is proportional to $(m_P^c - m_P)$ in the Gribov theory.
When $m_P$ is increased above $m_P^c$
these Goldstone states condense in the vacuum leading to a further
increase
in the mass of the constituent quark [2].
The quark mass which appears in the linear mass expansion (which does not
include ``pion" corrections)
for the baryon octet is the running mass $m_P$.
This means that $m_P$ is the light quark mass in equ.(4).
The phase transitions in the light quark vacuum mean that the Gribov
theory
anticipates a ``discrepancy" between the value of $\sigma_{\pi N}$ which is
measured in $\pi$ $N$ scattering
and the value of $\sigma_{\pi N}^{th}$ which is extracted from baryon
spectroscopy.
This ``discrepancy" would remain even if the effect of ``pion" corrections
to $M_B$ were included.
There is no theoretical need
to introduce a large strange quark matrix element in the nucleon in order to
explain the $\sigma_{\pi N}$ data.

We can use equ.(4) to estimate the value of $m_P^c$ at 1 GeV$^2$.
The old
spectroscopy prediction of the sigma term used $m_q = (m_P^c - m_P)$ instead
of $m_q = m_P$.
We substitute the measured value $\sigma_{\pi N} = 45$MeV into the linear
mass formula
$$
\sigma_{\pi N} ({m_s \over m_q} - 1) = constant
\eqno(9)
$$
and use the result that the strange ``current-quark" mass $m_s \simeq m_{s,P}$
to obtain $m_q \simeq 10.5$MeV at a scale $\mu^2 = 1$GeV$^2$.
It follows that the value of the critical mass for the chiral phase transition
is $m_P^c \simeq 16.5$MeV after evolution to $\mu^2 = 1$GeV$^2$.
A detailed analysis of pion and kaon physics in the Gribov theory will be
given in the next paper [9].

\pagebreak

{\bf REFERENCES}
\vskip 12pt
\begin{enumerate}
\item
R. L. Jaffe and C. L. Korpa, Comments Nucl. Part. Phys. 17 (1987) 163
\item
V. N. Gribov, Physica Scripta T15 (1987) 164, \\
Lund preprint LU TP 91/7 (May 1991) \\
Orsay lectures LPTHE Orsay 92/60 (June 1993)
\item
F. E. Close et al., Phys. Letts. B319 (1993) 291 \\
Yu. L. Dokshitzer, Gatchina preprint 1962 (1994)
\item
T. E. O. Ericson and W. Weise, {\it Pions and Nuclei}, chapter 9,
Oxford UP (1988) \\
A. W. Thomas, Adv. Nucl. Phys. 13 (1984) 1
\item
J. Gasser, H. Leutwyler and M. E. Sainio, Phys. Letts. B253 (1991) 251
\item
J. Gasser and H. Leutwyler, Phys. Reports 87 (1982) 77
\item
A. S. Gorsky, B. L. Ioffe and A. Yu. Khodjamirian, Phys. Letts. B227 (1989) 474
\item
V. N. Gribov, Nucl. Phys. B106 (1982) 103
\item
S. D. Bass and M. M. Taylor, in preparation
\end{enumerate}
\pagebreak
\end{document}